\documentclass[12pt]{article}  
\usepackage{cite}
\usepackage{epsfig}
\usepackage{graphicx}
\usepackage{amsmath}
\usepackage{amssymb}
\usepackage{ulem}
\usepackage{mdwlist}          
\usepackage{color}              

\usepackage{a41}
\usepackage{color}
\usepackage[rflt]{floatflt}
\usepackage{float}
\usepackage{slashed}

\usepackage{array}

\usepackage[english]{babel}
\usepackage[latin1]{inputenc}
\usepackage[T1]{fontenc}
\usepackage{ae}

\usepackage{url}

\usepackage{amsmath, amsthm, amssymb}
\newtheorem{thm}{Theorem}[section]

\newtheorem{definition}[thm]{Definition}

\newcommand{\gsim}{\raisebox{-0.07cm   }
{$\, \stackrel{>}{{\scriptstyle\sim}}\, $}}
 \newcommand{\GeV}{\mathrm{GeV}}

\newcommand{\ep}{\varepsilon}

\usepackage{rotating}

\usepackage{graphicx}

\newcounter{mmacnt}
\def\restartmma{\setcounter{mmacnt}{0}}
\restartmma \catcode`|=\active
\def|#1|{\mathrm{#1}}
\catcode`|=12
\newenvironment{mma}{
 \par\smallskip
 \catcode`|=\active
 \parskip=0pt\parindent=0pt 
 \small
 \def\In##1\\{%
   \def\linebreak{\hfill\break\null\qquad}%
   \refstepcounter{mmacnt}
   \hangindent=2.5em\hangafter=0
   \leavevmode
   \llap{\tiny\sffamily In[\arabic{mmacnt}]:=\kern.5em}%
   \mathversion{bold}\footnotesize$\displaystyle##1$\normalsize
   \mathversion{normal}\par
 }%
 \def\Print##1\\{%
   \def\linebreak{		\hfill\break}%
   \hangindent=2.5em\hangafter=0
   \leavevmode ##1\par}%
 \def\Out##1\\{%
   \def\linebreak{$\hfill\break\null\hfill$}%
   \kern\abovedisplayskip\par
   \hangindent=2.5em\hangafter=0
   \leavevmode
   \llap{\tiny\sffamily Out[\arabic{mmacnt}]=\kern.5em}
   \footnotesize$\displaystyle##1$\normalsize\hfill\null\par
   \kern\belowdisplayskip
 }%
 \def\Warning##1##2\\{%
   \def\linebreak{\hfill\break}%
   \hangindent=2.5em\hangafter=0
   \leavevmode
   {\scriptsize##1 : ##2}\par}%
}{%
 \par\smallskip
}

\usepackage{color}

\newenvironment{fshaded}{%
\MakeFramed {\FrameRestore}
}%
{\endMakeFramed}

\allowdisplaybreaks[4]

\begin{document}
\setlength{\baselineskip}{0.515cm}
\sloppy
\thispagestyle{empty}
\begin{flushleft}
DESY 24--066
\hfill 
\end{flushleft}

\mbox{}
\vspace*{\fill}
\begin{center}

{\LARGE\bf Next-to-Next-to-Leading Order Evolution of}

\vspace*{3mm} 
{\LARGE\bf  Polarized Parton Densities in the Larin Scheme} 

\vspace{3cm}
\large
J.~Bl\"umlein$^{a,b}$ and M.~Saragnese$^{a,c}$

\vspace{1.cm}
\normalsize

\vspace*{2mm}
{\it $^a$ Deutsches Elektronen-Synchrotron DESY, Platanenallee 6, 15738 Zeuthen, Germany} 

\vspace*{2mm}
{\it $^b$ Institut f\"ur Theoretische Physik III, IV, TU Dortmund, Otto-Hahn Stra\ss{}e 4, \newline 44227 Dortmund, 
Germany}

\vspace*{2mm}
{\it $^c$ 
Wolfram Research, Inc., 100 Trade Center Drive Champaign, IL 61820-7237, USA}

\vspace*{3mm}


\end{center}
\normalsize
\vspace{\fill}
\begin{abstract}
\noindent
In many calculations involving polarized twist-2 parton densities to higher order in the strong coupling
constant one uses the Larin scheme to describe chiral effects in dimensional regularization. Upon forming
observables, the scheme dependence cancels. Still one needs a corresponding regularization scheme to
compute the contributing building blocks, like massless and massive Wilson coefficients, as well 
as the massive 3-loop operator matrix elements used in the variable flavor number scheme. These are matched 
to the evolved parton distribution functions in the Larin scheme. Starting with suitable input distributions
we provide the solution of scale evolution of the different polarized parton distribution functions in 
Bjorken $x$ space for a wide range of virtualities $Q^2$ in the Larin scheme, at  next-to-leading, 
and to next-to-next-to-leading order for the first time. We also illustrate the deviation between the parton 
distributions in the Larin and $\overline{\rm MS}$ schemes numerically.
\end{abstract}

\vspace*{\fill}
\noindent

\newpage

\vspace*{1mm}
\noindent
\section{Introduction}
\label{sec:1}

\vspace*{1mm}
\noindent
In the calculation of the scaling violations of polarized hadronic processes the building blocks of the 
factorized processes are scheme dependent. Working in dimensional regularization one may either use 
the Larin scheme \cite{Larin:1993tq} or the HVBM scheme \cite{tHooft:1972tcz,Akyeampong:1973xi,
Akyeampong:1973vk,Akyeampong:1973vj,Breitenlohner:1975hg,Breitenlohner:1976te} to describe the Dirac 
matrix $\gamma_5$ in $D = 4 + \ep$ dimensions. Computing the Lorentz- and Dirac-structures of the 
corresponding scattering processes by using {\tt Form} \cite{Vermaseren:2000nd}, reference to the 
Larin--scheme is particularly efficient. Finally, one forms observables, such as the structure functions
of deep-inelastic scattering \cite{Politzer:1974fr,Geyer:1977gv,Buras:1979yt,Reya:1979zk,Lampe:1998eu,
Blumlein:2012bf}, or other hard-scattering processes. The initial scheme dependence of the Wilson 
coefficients and parton distribution functions (PDFs) cancels in the observables.\footnote{
Alternatively, one may consider scheme-invariant evolution of structure functions, 
see~Refs.~\cite{Blumlein:2004xs,Blumlein:2021lmf} and references therein.}

For this reason, we provide the solution of the polarized partonic evolution equations in the Larin 
scheme 
at  next-to-leading order (NLO) and 
next-to-next-to-leading order (NNLO) in the present paper, 
following earlier representations at NLO in the $\overline{\rm MS}$
scheme \cite{Blumlein:2010rn}. At leading order (LO) the evolution of parton distributions is scheme-invariant.
To perform a consistent expansion of the corresponding evolution equations 
algebraically we work in Mellin--$N$ space, cf.~Ref.~\cite{Blumlein:1997em}. The results are 
presented in terms of grids in the virtuality $Q^2$ and the Bjorken variable $x$ for $4~\GeV^2 \leq Q^2 
\leq 10^6~\GeV^2$ and  $10^{-9} \leq x < 1$.\footnote{Lower values of $Q^2$ lay outside the deep--inelastic region 
and require the separation of resonant and (quasi)elastic contributions.}
The shapes 
of the PDFs are parameterized at 
$Q^2_0 = 4~\GeV^2$, fixing $\alpha_s^{\rm NLO}(M_Z^2)$ = 0.1191 from the fit \cite{Alekhin:2018pai} 
and $\alpha_s^{\rm NNLO}(M_Z^2)$ = 0.1147 from the analysis in Ref.~\cite{Alekhin:2017kpj}. 
Here the strong coupling constant is defined by $\alpha_s = g_{s,\rm ren}^2/(4\pi) \equiv a_s (4 
\pi)$. In the leading order evolution we use the starting  value $\alpha_s^{\rm NLO}(Q_0^2)$, because 
leading order QCD--fits are unstable. This leads to $\alpha_s^{\rm LO}(M_Z^2)$ = 0.1247. The values of 
$\alpha_s(M_Z^2)$ determined in the polarized data analysis \cite{Blumlein:2010rn} are consistent with 
these values, although the current errors are much larger than for the values obtained using unpolarized 
deep--inelastic scattering data due to the current statistics of the polarized World data. 

The polarized anomalous dimensions at LO were calculated in \cite{Sasaki:1975hk,Ahmed:1975tj,
Altarelli:1977zs}, at NLO in \cite{Mertig:1995ny,Vogelsang:1995vh,Vogelsang:1996im} and NNLO in 
Refs.~\cite{Moch:2004pa,Moch:2014sna,Behring:2019tus,Blumlein:2021enk,Blumlein:2021ryt}.
The massless Wilson coefficients at $O(a_s)$ were computed in \cite{Furmanski:1981cw,Kodaira:1978sh,
Kodaira:1979ib,Antoniadis:1980dg,Bodwin:1989nz}, at $O(a_s^2)$ in \cite{Zijlstra:1993sh,Vogt:2008yw} 
and at $O(a_s^3)$ in the non-singlet case in \cite{Vermaseren:2005qc,Blumlein:2022gpp} and the singlet case 
in
Ref.~\cite{Blumlein:2022gpp}. Finally, the massive Wilson coefficients at LO 
were obtained in \cite{Watson:1981ce}. Numerical results at NLO were computed in \cite{Hekhorn:2018ywm}. 
In the asymptotic region $Q^2 \gg m_Q^2$ the massive Wilson coefficients and massive operator matrix 
elements (OMEs) were calculated at NLO in \cite{Buza:1996xr,Bierenbaum:2007pn,
Blumlein:2019zux,Bierenbaum:2022biv}, even before \cite{Hekhorn:2018ywm}, and at NNLO in 
Refs.~\cite{Ablinger:2014vwa,Behring:2015zaa,Ablinger:2019etw,Behring:2021asx,Blumlein:2021xlc,
Ablinger:2022wbb,Ablinger:2023ahe, Ablinger:2024xtt} in the single mass case. The corresponding 
3--loop two--mass corrections were calculated in Refs.~\cite{Ablinger:2017err,Ablinger:2019gpu,
Ablinger:2020snj}. For all quantities up to $O(a_s^2)$ the transformation from the Larin to the 
$\overline{\rm MS}$ scheme is known. The polarized asymptotic heavy flavor contributions of $O(a_s^3)$, 
i.e. the NNLO corrections, depend also on the 3--loop massless Wilson coefficients and are given in 
the Larin scheme. This also applies to the transition coefficients in the variable flavor number 
scheme at $O(a_s^3)$. In these cases the PDFs in the Larin scheme have to be used in representing 
the scaling violations of the corresponding observables. It finally turns out that the deviations of the 
evolution of different parton densities are larger than the projected accuracy for future polarized 
deep-inelastic scattering experiments of $O(1\%)$ \cite{AbdulKhalek:2021gbh}.

The paper is organized as follows. In Section~\ref{sec:2} we describe the parameterization of the 
polarized parton densities and details of the evolution. Some numerical comparisons of the results in 
the $\overline{\rm MS}$ and Larin schemes are given in Section~\ref{sec:3}, and Section~\ref{sec:4} 
contains the conclusions. The parameterizations in the $\overline{\rm MS}$ and Larin schemes are 
provided in ancillary files in form of grids and a numerical program which allows for spline 
interpolation in the kinematic regions considered.
\section{The Parameterization and Details of the Evolution}
\label{sec:2}

\vspace*{1mm}
\noindent
The PDFs of the polarized quarks and antiquarks $\Delta q_i(x,Q^2)$, as well as of the polarized gluon 
density $\Delta G(x,Q^2)$, are parameterized at the scale $Q^2_0 =  4~\GeV^2$, 
cf.~\cite{Blumlein:2010rn}\footnote{Here it has been shown, that this parameterization 
is close to those obtained in other analyses, see e.g.~\cite{Goto:1999by,Hirai:2003pm,Hirai:2006sr,Hirai:2008aj, 
Gluck:2000dy,Leader:2001kh,deFlorian:2008mr,deFlorian:2009vb} and Figures~2 and 3, \cite{Blumlein:2010rn}, for
comparisons at the initial scale $Q_0^2$.}, by
\begin{eqnarray}
\label{param1}
x\Delta f_i(x,Q_0^2) = \eta_i A_i x^{a_i} (1 - x)^{b_i}
\left(1 + \gamma_i x\right),~~~~f = q, G
\end{eqnarray}
with the normalization constants $A_i$, being given by
\begin{eqnarray}
\label{anorm1}
A_i^{-1} & = & \left( 1 + \gamma_i\frac{a_i}{a_i+b_i+1} \right)
               B(a_i,b_i+1).
\end{eqnarray}
The parameters $\eta_i$ denote the first moments
\begin{eqnarray}
\eta_i = \int_0^1dx \Delta f_i(x,Q_0^2)
\end{eqnarray}
of the respective distributions. The values of the initial parameters are given in Table~2 of 
Ref.~\cite{Blumlein:2010rn}. Here $\eta_{u_v}$ and $\eta_{d_v}$ are fixed by the $F$ and $D$ 
parameters measured in neutron and hyperon $\beta$--decays. $B(a,b) = \Gamma(a) 
\Gamma(b)/\Gamma(a+b)$ denotes Euler's $B$--function, related to Euler's $\Gamma$-function.

In the case of three massless flavors the following two flavor non--singlet distributions contribute
\begin{eqnarray}
\label{eq:NS3}
\Delta_3^{\rm NS,-} &=&
  (\Delta u + \Delta \bar{u}) - (\Delta d + \Delta \bar{d}),
\\
\label{eq:NS8}
\Delta_8^{\rm NS,-} &=& 
  (\Delta u + \Delta \bar{u}) + (\Delta d + \Delta \bar{d})
- 2 (\Delta s + \Delta \bar{s}).
\end{eqnarray}
Furthermore, there is the non--singlet distribution $\Delta q^{\rm NS,v}$,
\begin{eqnarray}
\Delta q^{\rm NS,v} =  \sum_{i=1}^3  \left[\Delta q_i - \Delta \bar{q}_i\right].
\label{eq:g5}
\end{eqnarray}
The distributions $\Delta_{3(8)}^{\rm NS,-}$ evolve with the splitting function $\Delta P^{\rm NS,-} 
= P^{\rm NS,+}$ and $q^{\rm NS,v}$ with $\Delta P^{\rm NS,-} + \Delta P^{\rm NS,s}$.
The splitting function  $\Delta P^{\rm NS,s}$ is computable at even values of $N$ and is continued 
to $N \in \mathbb{C}$.\footnote{Our evolution programme
is written such, that all quantities are free of factors $(-1)^N$. This can be obtained as outlined in
Refs.~\cite{Blumlein:1998if,Blumlein:2006mh}.}
$\Delta P^{\rm NS,s}$ \cite{Moch:2015usa,Blumlein:2021ryt}
is scheme--independent at three--loop order and $P^{\rm NS,+}$ derives from a vector current. 
Therefore, to three--loop order, there is no impact of the Larin scheme on this quantity.
The distribution (\ref{eq:g5}) can be measured from the interference structure function $g_5^-(x,Q^2)$, see 
Ref.~\cite{Blumlein:1996vs}. 
For a completely symmetric polarized sea, $\Delta {u}_s = \Delta {d}_s =  
\Delta \bar{u} = \Delta \bar{d} = \Delta s = \Delta \bar{s}$,
which we will consider, the relation
\begin{eqnarray}
\label{eq:NSv}
\Delta q^{\rm NS,v}(x) =  \Delta_8^{\rm NS,-}(x)
\end{eqnarray}
holds at the input scale $Q_0^2$. 

The polarized singlet distribution is given by 
\begin{eqnarray}
\Delta \Sigma = \sum_{i=1}^3 \left[\Delta q_i + \Delta \bar{q}_i\right], 
\end{eqnarray}
which evolves together with the polarized gluon distribution $\Delta G$. The differences in the 
anomalous dimensions between the Larin and the $\overline{\rm MS}$ scheme start at NLO. They are implied
by the correction terms $z_{qq}^{(k), \rm NS}$ and $z_{qq}^{(k+1), \rm PS},~~k \geq 1$, 
cf.~\cite{Matiounine:1998re,Moch:2014sna,Blumlein:2021ryt}.

The evolution equations in the polarized case consist of three flavor non--singlet evolution equations to NNLO
\cite{Moch:2004pa,Blumlein:2021enk} and the coupled singlet and gluon distribution evolution 
\cite{Moch:2014sna,Blumlein:2021ryt}. In the latter case one solves matrix-valued differential equations.
In Mellin--$N$ space the corresponding iterative solutions can be obtained analytically. Here one expands 
systematically in $a_s(\mu^2)$, with $\mu^2 = Q^2$ and $\mu^2 = Q_0^2$ up to a given order $l$ with
$O(a_s^{l-m}(Q^2) a_s^m(Q_0^2)), m \in \{0, ...,l\}$, see
Ref.~\cite{Blumlein:1997em}, Section~5.1, for details. The solution in $x$--space is obtained by 
a single contour integral around the singularities of the problem, see~\cite{Blumlein:2000hw,BS}.
This representation avoids pile--up contributions for the parton distributions by iterative solutions 
of the evolution equation, which were discussed in Ref.~\cite{Blumlein:1996gv}. They have, otherwise, 
to be removed by using the method described in \cite{Rossi:1983bp,Rossi:1983xz}, see also 
\cite{Gluck:1991ee}.\footnote{One may even expand 
whole observables in $a_s(Q^2)$ and $a_s(Q_0^2)$ to obtain completely scheme-invariant quantities, see 
Ref.~\cite{Blumlein:2021lmf}.}
The evolution of the strong coupling constant $a_s(Q^2)$ is calculated numerically in the respective regions
of constant $N_F$, cf. e.g.~\cite{Vogt:2004ns}, performing the matching at $Q^2=m_c^2$, $m_b^2$, and $m_t^2$ 
by using the on--shell masses \cite{Alekhin:2012vu,ParticleDataGroup:2020ssz},
\begin{eqnarray}
m_c = 1.59~\GeV,~~~~~~~~~~~
m_b = 4.78~\GeV,~~~~~~~~~~~
m_t = 172.5~\GeV,
\end{eqnarray}
according to \cite{Chetyrkin:1997sg}. The expansion coefficients of the $\beta$--function were calculated in 
Refs.~\cite{Gross:1973id,Politzer:1973fx,Caswell:1974gg,Jones:1974mm,Tarasov:1980au,Larin:1993tp,
vanRitbergen:1997va,Czakon:2004bu,Chetyrkin:2004mf,Baikov:2016tgj,Herzog:2017ohr,Luthe:2017ttg}.
\section{Numerical Results}
\label{sec:3}

\vspace*{1mm}
\noindent
In the following we consider the evolution of the non--singlet distributions (\ref{eq:NS3}, 
\ref{eq:NS8}) and (\ref{eq:g5}) and of the polarized singlet and gluon distribution.
Depending on whether the structure function $g_1(x,Q^2)$ is measured at a proton or deuteron target
(after nuclear corrections), cf.~\cite{Blumlein:2021lmf}, both the distributions (\ref{eq:NS3}, 
\ref{eq:NS8}) or only (\ref{eq:NS8}) contribute, while (\ref{eq:g5}) appears only for polarized 
structure functions with an axialvector--vector contribution, as e.g. for $g_5^-(x,Q^2)$.
To three--loop order the choice of the scheme has, however, no effect on the evolution of this 
distribution.
For this reason we do not consider this case here. 

The grid files are named {\tt PolLO(NLO,NNLO)M(L)} for the PDFs contributing to $g_1(x,Q^2)$, respectively. 
Here, the files at LO are the 
same in the Larin (L) and $\overline{\rm MS}$ (M) scheme. We use a cubic spline 
interpolation\footnote{We thank S. Kumano and M. Miyama of the AAC-collaboration for allowing us to 
use their interpolation routines.} both in $x$ and $Q^2$ in the parameterized kinematic region, 
see~Section~\ref{sec:1}. The attached {\tt FORTRAN} code {\tt Npolpdf.f} provides the values of the 
respective distributions.

\begin{figure}[H]
\centering
\includegraphics[width=0.49\textwidth]{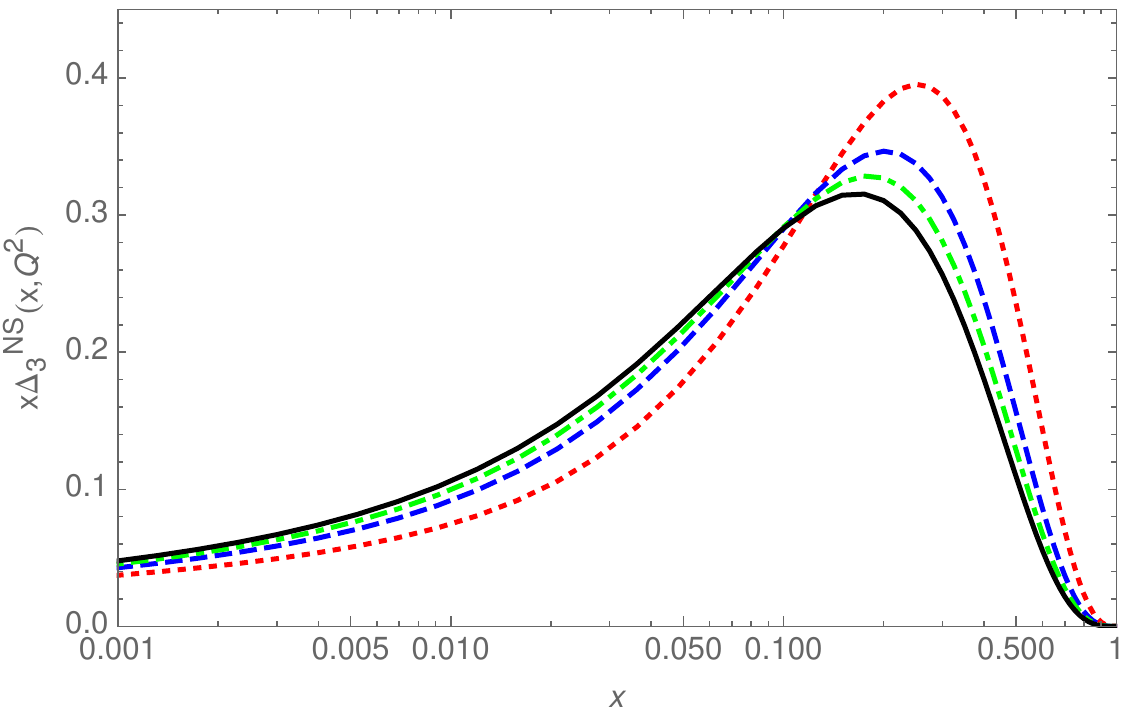}
\includegraphics[width=0.49\textwidth]{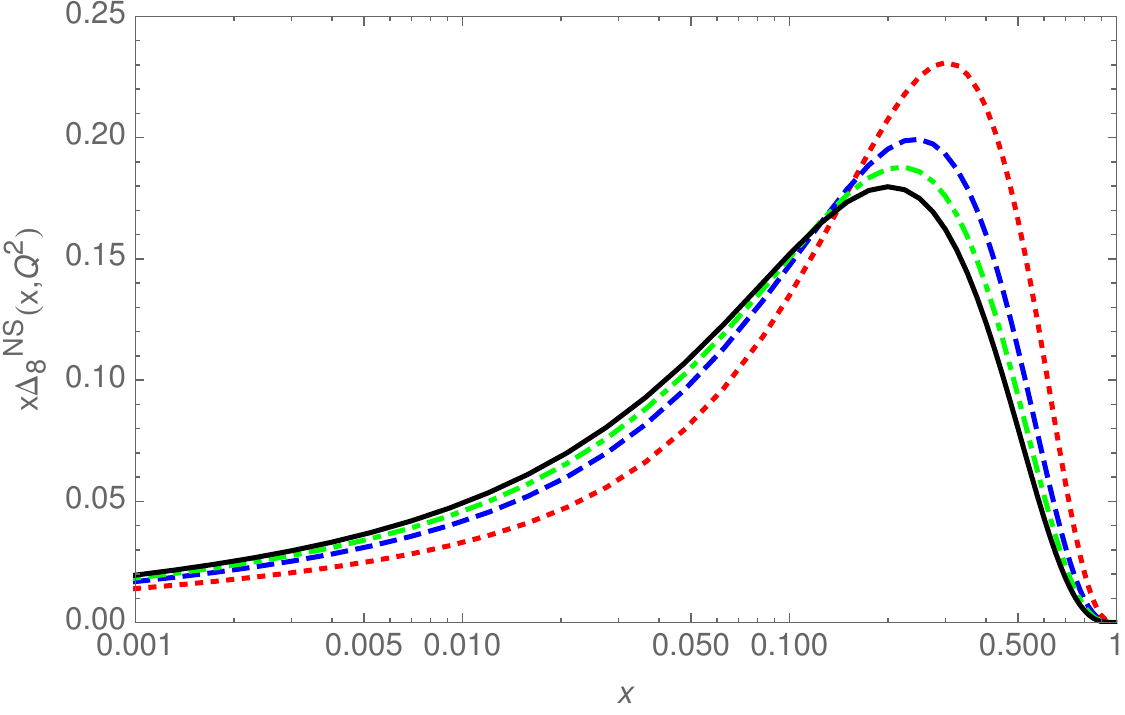}
\caption{\sf The evolution of the polarized non--singlet densities
$x\Delta_3^{\rm NS,-}$ and $x\Delta_8^{\rm NS,-}$ in the $\overline{\rm MS}$ scheme at NNLO.
Dotted lines: $Q^2 = 4~\GeV^2$,
dashed lines: $Q^2 = 100~\GeV^2$,
dash-dotted lines: $Q^2 = 1000~\GeV^2$,
full lines: $Q^2 = 10000~\GeV^2$.
\label{fig:1}}
\end{figure}

\begin{figure}[H]
\centering
\includegraphics[width=0.49\textwidth]{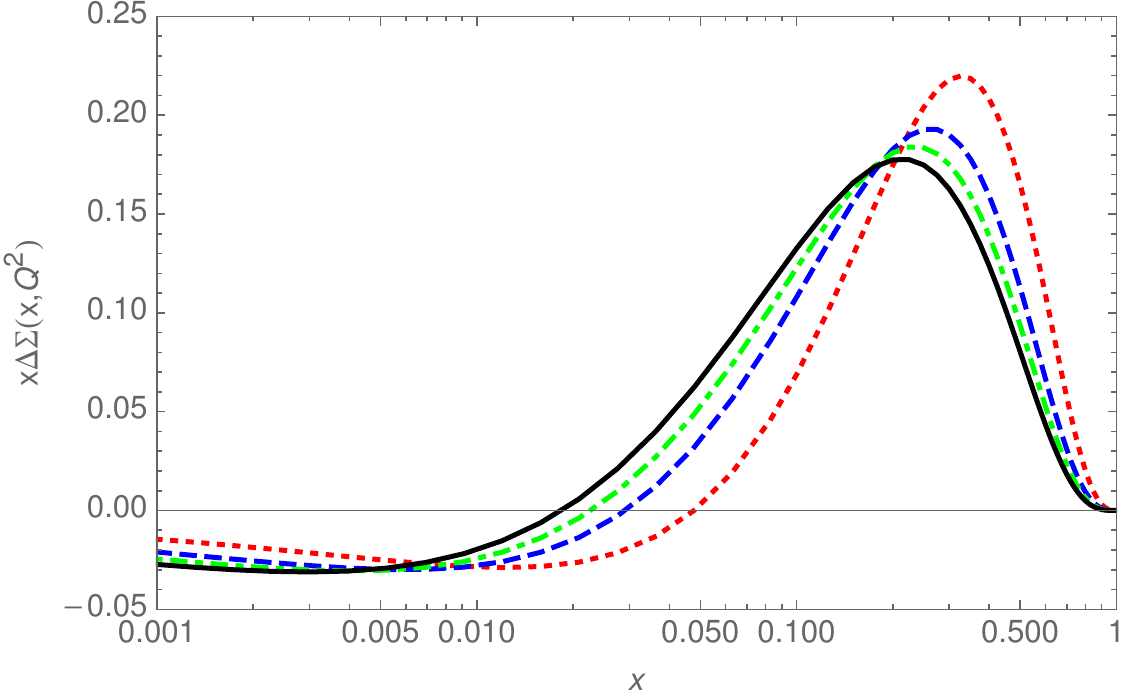}
\includegraphics[width=0.49\textwidth]{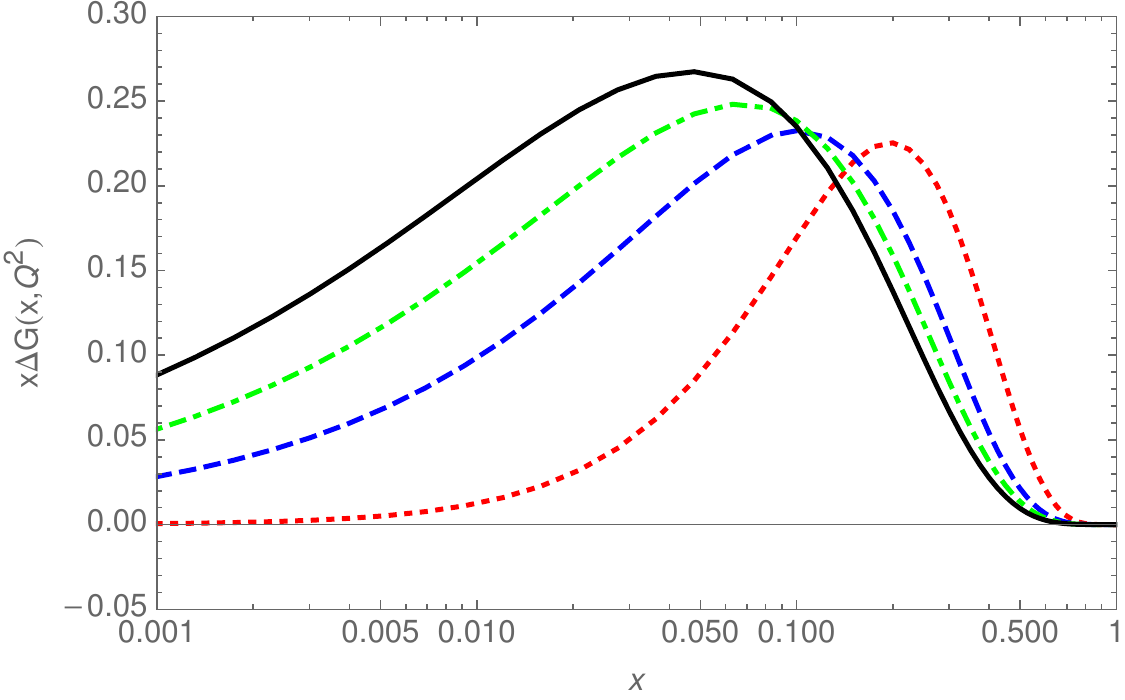}
\caption{\sf The evolution of the polarized singlet and gluon densities
$x\Delta \Sigma$ and $x\Delta G$ in the $\overline{\rm MS}$ scheme at NNLO.
Dotted lines: $Q^2 = 4~\GeV^2$,
dashed lines: $Q^2 = 100~\GeV^2$,
dash-dotted lines: $Q^2 = 1000~\GeV^2$,
full lines: $Q^2 = 10000~\GeV^2$.
\label{fig:2}}
\end{figure}

We first illustrate the evolution of the $x\Delta_3^{\rm NS,-}, x\Delta_8^{\rm NS,-}, x\Delta \Sigma$, and 
$x\Delta G$, as functions in $x$ and $Q^2$ in the $\overline{\rm MS}$ scheme in 
Figures~\ref{fig:1}--\ref{fig:2} at NNLO. 
While the maximum of the distributions deplete in the quarkonic case with
growing values of $Q^2$, they rise for the gluon. In the gluonic case the maxima shift more strongly 
to lower values of $x$, compared to the quarkonic cases. The evolution leads to a growth of the distributions
in the range of smaller values of $x$, while they are depleted in the large $x$ region for rising values of the
virtuality $Q^2$. The singlet distribution is negative in the region of smaller values of $x$ due to the negative sea quark 
distributions.

\begin{figure}[H]
\centering
\includegraphics[width=0.49\textwidth]{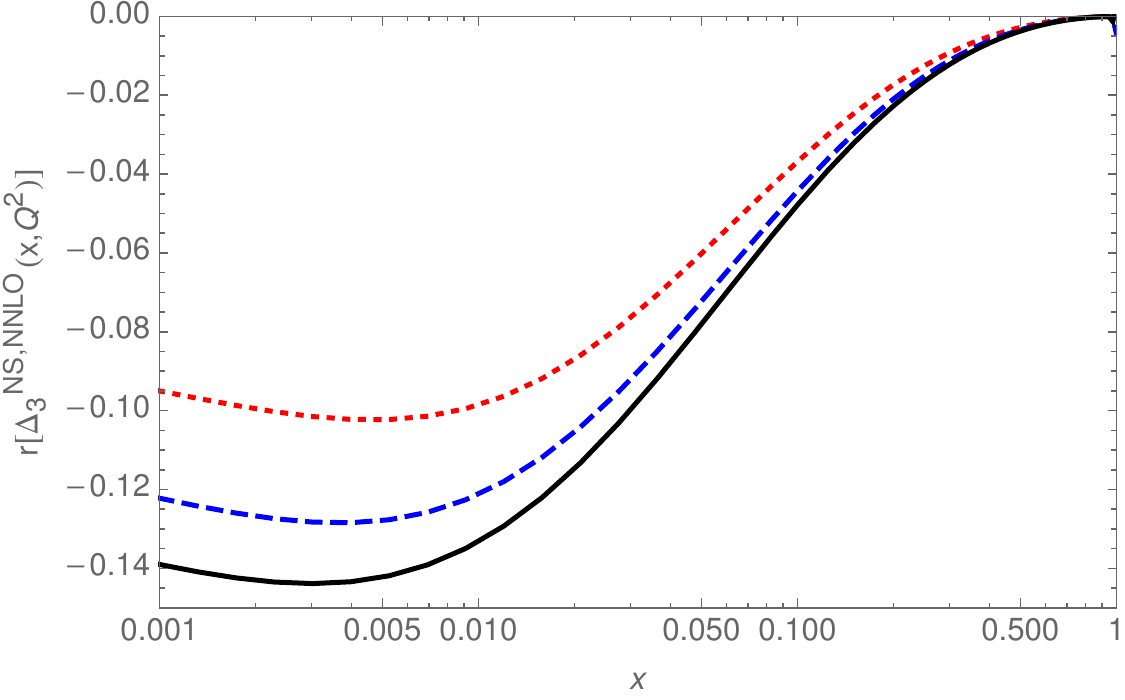}
\includegraphics[width=0.49\textwidth]{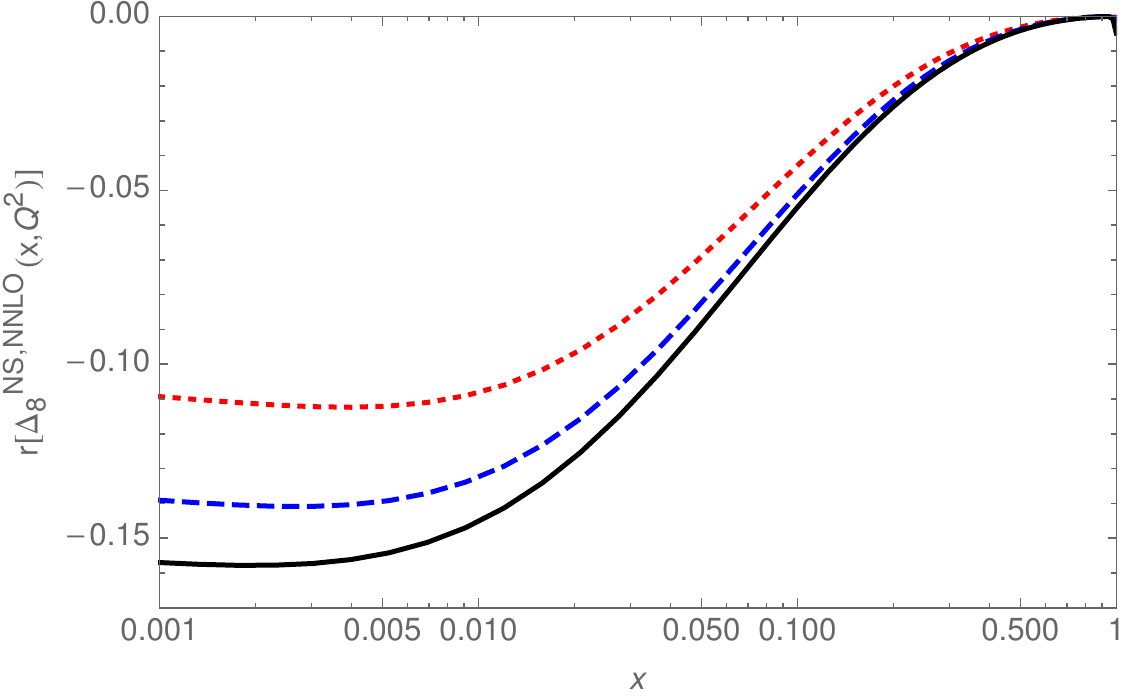}
\caption{\sf The function $r(x,Q^2)$, Eq.~(\ref{eq:rat}), for the evolution at NNLO; 
dotted lines: $Q^2 = 100~\GeV^2$,
dashed lines: $Q^2 = 1000~\GeV^2$,
full lines:   $Q^2 = 10000~\GeV^2$,
Left panel: $\Delta_3^{\rm NS,-}$. 
Right panel: $\Delta_8^{\rm NS,-}$. 
\label{fig:4}}
\end{figure}

We now compare the deviations of the different distributions considered in the Larin (L) and 
$\overline{\rm MS}$ scheme (M), 
\begin{eqnarray}
\label{eq:rat}
r(x,Q^2) = \frac{f^{\rm L}(x,Q^2)}{f^{\rm M}(x,Q^2)} - 1
\end{eqnarray}
at NNLO for the values of $Q^2 = 100, 1000$ and $10000~\GeV^2$ as functions  of $x$. 
In Figure~\ref{fig:4} we illustrate the deviation for the non--singlet distributions 
$x\Delta_3^{\rm NS,-}$ and $x\Delta_8^{\rm NS,-}$. The deviations are similar for both distributions
and grow towards small value of $x$.

\begin{figure}[H]
\centering
\includegraphics[width=0.49\textwidth]{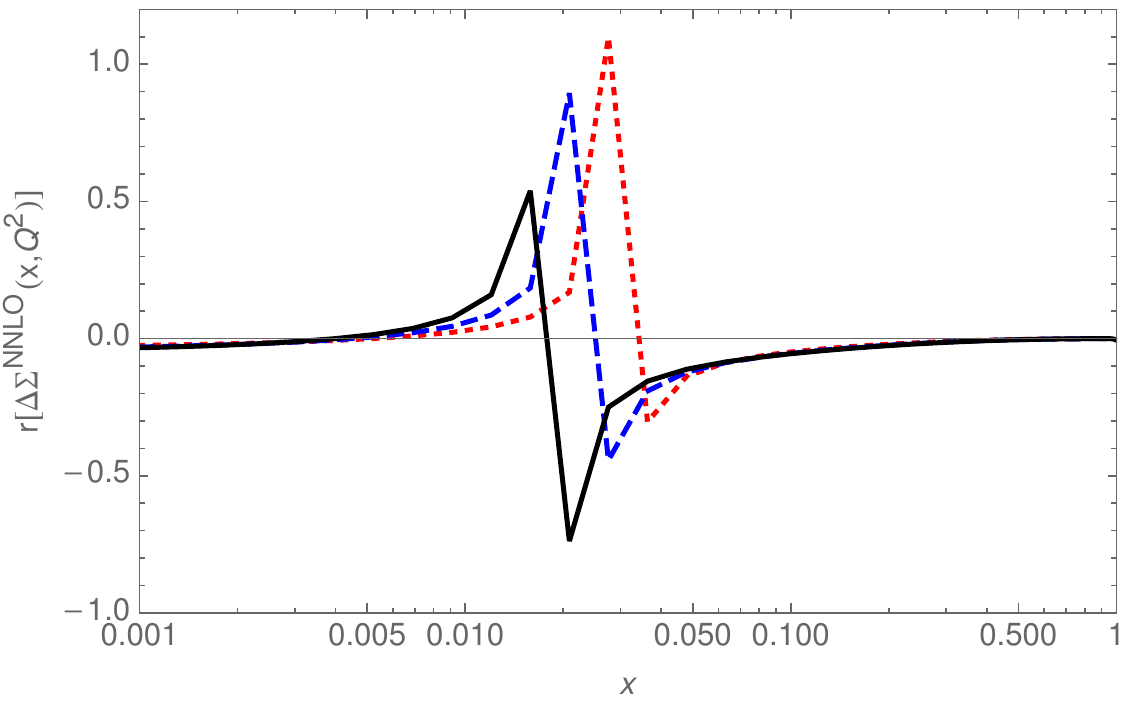}
\includegraphics[width=0.49\textwidth]{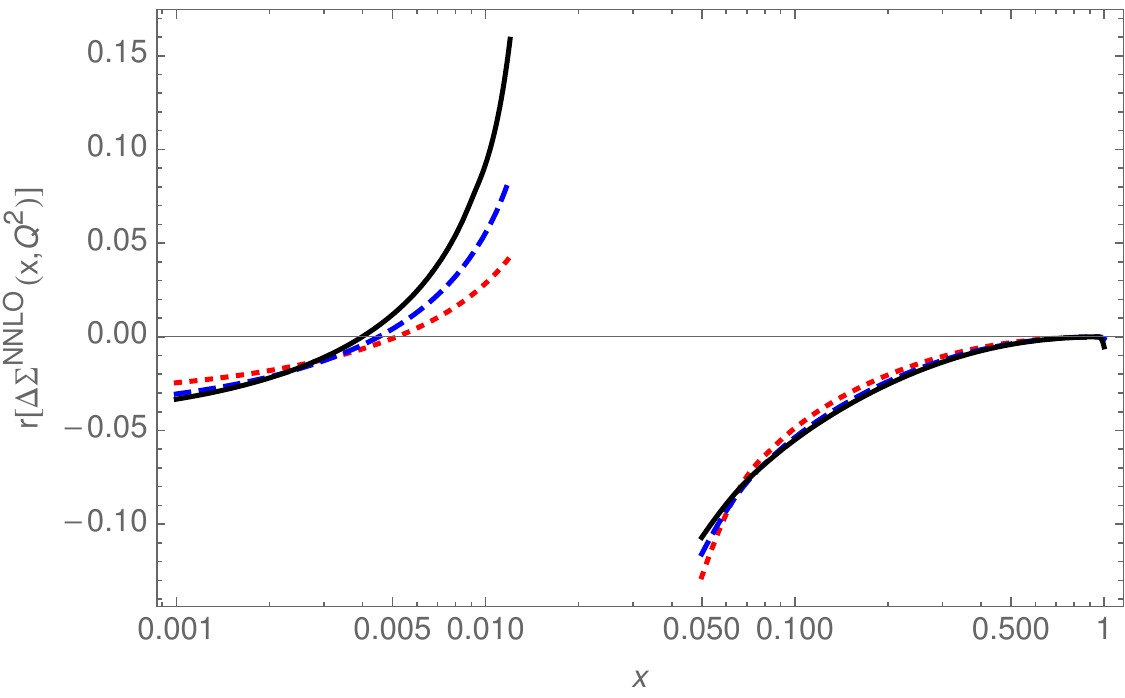}
\caption{\sf The function $r(x,Q^2)$, Eq.~(\ref{eq:rat}), for the evolution at NNLO;
dotted lines: $Q^2 = 100~\GeV^2$,
dashed lines: $Q^2 = 1000~\GeV^2$,
full lines:   $Q^2 = 10000~\GeV^2$,
Left panel: $\Delta \Sigma$. 
Right panel: $\Delta \Sigma$, enlarging the small and large $x$ regions and  excluding the zero-transition 
region.
\label{fig:5}}
\end{figure}

\begin{figure}[H]
\centering
\includegraphics[width=0.49\textwidth]{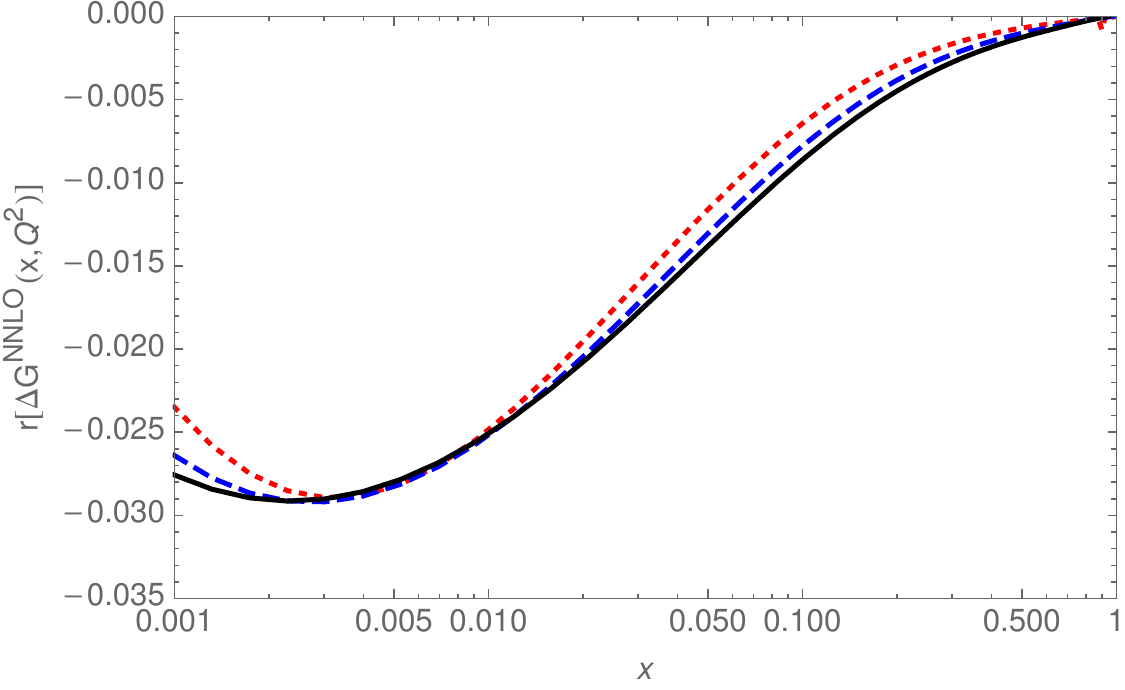}
\caption{\sf The function $r(x,Q^2)$, Eq.~(\ref{eq:rat}) for the evolution of $\Delta G$ at NNLO;
dotted lines: $Q^2 = 100~\GeV^2$,
dashed lines: $Q^2 = 1000~\GeV^2$,
full lines:   $Q^2 = 10000~\GeV^2$.
\label{fig:5a}}
\end{figure}

In Figures~\ref{fig:5} and \ref{fig:5a} we show the deviation for the singlet 
distributions $x\Delta \Sigma$ and $x\Delta G$. While the effect is larger for 
$x\Delta \Sigma$ due to the change in the quarkonic splitting functions, the one for
$x\Delta G$ is relatively small, since the splitting functions $\Delta P_{gg}^{(k)}$ are not affected 
and the changes come only from convolutions with other splitting functions. 
The large relative effects in the range of medium values of $x$ in the ratio 
for $x\Delta \Sigma$ are due to zero-transitions for this quantity. Similar results are 
obtained at NLO. 

\noindent
All ratios $r(x,Q^2)$ turn to zero for $x \rightarrow 1$. This is due to the fact that the difference of splitting 
functions $\Delta P_{qq}^{+,(1),\rm NS}, \Delta P_{qg}^{(1)}$ and $\Delta P_{gq}^{(1)}$ are $\propto 1/N^3$ and
$\Delta P_{qq}^{+,(2),\rm NS} \propto \ln(N)/N^3$, $\Delta P_{qq}^{(2),\rm PS} \propto 1/N^3$ and 
$\Delta P_{qg}^{(2)}, \Delta P_{gq}^{(2)} \propto \ln^2(N)/N^3$, where the large $N$ behaviour rules 
that at large values of $x$.  
From Figures~\ref{fig:4}--\ref{fig:5} one sees that the evolution of the polarized parton densities are 
different
in the $\overline{\rm MS}$  and the Larin schemes, although at different degree. They have to be taken into 
account in precision analyses.
\section{Conclusions}
\label{sec:4}

\vspace*{1mm}
\noindent
We have calculated the evolution of polarized parton distribution functions in the Larin scheme in a 
wide range of the kinematic variables $x$ and $Q^2$ at NLO and NNLO. These distributions can be used
to form observables in cases in which polarized Wilson coefficients, hard-scattering sub-system cross 
sections, or massive operator matrix elements were calculated in this scheme only. This allows for a 
consistent data analysis. It is already clear from the results on the non--singlet distributions leading to
corrections of up to $O(15\%)$ in the small $x$ range, $x \gsim 0.001$, that the polarized parton distributions 
in both schemes are significantly different, given e.g. the future experimental accuracies to be reached at the 
EIC \cite{AbdulKhalek:2021gbh,Boer:2011fh}. One therefore needs the PDFs in the Larin scheme, to describe the 
flavor matching in 
the variable flavor number scheme and for incorporating the heavy flavor Wilson coefficients in the polarized case
into data analysis of deep--inelastic structure functions and hard scattering cross sections at polarized 
hadron 
colliders. 

For comparison, we also performed the corresponding evolution in the $\overline{\rm MS}$ scheme at LO, NLO and 
NNLO. We provide both sets of grids and the corresponding {\tt FORTRAN} program in an ancillary file to this paper.

\vspace*{5mm}
\noindent
{\bf Acknowledgment.} 
We would like to thank A.~Behring, P.~Marquard, E.~Reya and  K.~Sch\"onwald for discussions. 

{\small

}
\end{document}